\documentclass[aps,prl,a4paper,reprint,superscriptaddress,showpacs]{revtex4-1}
\usepackage{graphicx,graphics,color,epsfig}
\usepackage{epstopdf} 
\usepackage{bm}
\usepackage{hyperref}
\usepackage{amsmath}
\usepackage{amsfonts}
\usepackage{amssymb}
\usepackage{appendix}
\usepackage{booktabs}

\begin{document}

\title{Magnetic Field-Free Giant Magnetoresistance in
a Proximity- and\\ Gate-Induced Graphene Spin Valve
}

\author{Yu Song}
\email{kwungyusung@gmail.com}
\affiliation{Microsystem and Terahertz Research Center, China Academy of Engineering Physics,
Chengdu 610200, P.R. China}
\affiliation{Institute of Electronic Engineering, China Academy of Engineering Physics,
Mianyang 621999, P.R. China}

\begin{abstract}
Due to its two dimensional nature, ferromagnetism and charge doping can be induced by proximity and electric field effects in graphene. Taking advantage of these features, we propose an electrically engineered spin valve by combining two magnetic insulators (using EuO, EuS, or YIG) and three coating gates. Two top gates are used to cancel the heavy electron doping's in these magnets and one back gate is used to utilize the normal or half-metallic ferromagnetisms. We demonstrate that, when the second top gate is tuned to utilize the insulating or spin insulating states, huge giant magnetoresistance (GMR) at high temperature (several times of $10^5\%$ at 68K and 100K) can be achieved for EuO and YIG. These results imply a distinguished GMR that is magnetism tunable, vertical configured (ferromagnetism versus insulating), and magnetic field-free. Our work may offer a viable path to a tantalizing magnetic field-free spintronics.\end{abstract}

\date{\today} 
\maketitle


Graphene, although a diamagnetic material,
is highly promising for spintronics.
This is because it supports not only long diffusion lengths and long spin lifetimes at room temperature,
but also magnetic moments induced by various methods \cite{han2014graphene}.
Introduction of vacancy defects,
doping with molecules or elements with high spin-orbital coupling,
and tailoring as zig-zag edged nanoribbons
can induce ferromagnetism in graphene \cite{han2014graphene,feng2017prospects}.

Among the proposed methods, graphene coupling with nearby magnetic insulators
are the most intriguing way \cite{feng2017prospects}.
Theoretically, EuO, EuS, and yttrium iron garnet Y$_3$Fe$_5$O$_{12}$ (YIG)
have been predicted to induce ferromagnetism with heavy electron doping
in graphene through
proximity effect \cite{haugen2008spin,yang2013proximity,hallal2017tailoring}.
Nontrivial effects, such as simultaneous spin filter and spin valve effect \cite{song2015spin},
pure crossed Andreev reflection \cite{ang2016nonlocal}, 
and quantum anomalous Hall effect \cite{su2017effect}
have been proposed in graphene-EuO heterostructures.
Experimentally, EuO has been integrated on graphene,
in which ferromagnetism with 67K Curie temperature
and heavy electron doping was confirmed \cite{swartz2012integration,swartz2013integrating}.
On the other hand, anomalous Hall effect \cite{wang2015proximity},
spin-current convention \cite{mendes2015spin},
spin transport \cite{leutenantsmeyer2016proximity},
and chiral charge pumping \cite{evelt2017chiral}
have been demonstrated as a probe of the ferromagnetism
in a graphene/YIG heterostructure.
Similarly, Zeeman spin Hall effect \cite{wei2016strong} was exhibited
for that in a graphene/EuS heterostructure.
These works reveal graphene on EuO, EuS, or YIG as emergency 2D ferromagnets.

\begin{figure}[t]
  \centering
  \includegraphics[width=0.95\linewidth]{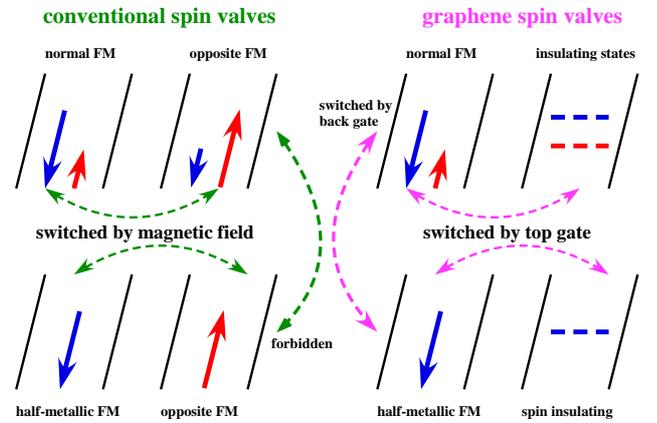}\\
  \caption{GMR mechanisms for conventional and graphene spin valves.
 In conventional spin valves (the left side),
 GMR can stem from normal \cite{baibich1988giant} or half-metallic  \cite{hwang1997enhanced} ferromagnetism.
  An anti-parallel (AP) configuration is responsible for the high resistance states
  and can be switched from the parallel (P) configuration by a magnetic field.
  In graphene spin valves (the right side),
  GMR can be supported by both normal and half-metallic ferromagnetisms.
  A vertical (V) configuration is responsible for the high resistance state,
  and can be switched from the P configuration by an electric field.
  The two magnetisms are also switched by a back gate.
}\label{mechanism}
\end{figure}

In this work, we 
explore giant magnetoresistance (GMR) applications
of these ferromagnets.
We propose a spin valve based on two magnetic insulators and three coating gates.
Of them, two top gates are used to cancel the electron doping 
through the strong electric field effect \cite{novoselov2004electric,kim2012direct},
and one back gate is used to
utilize a normal or half-metallic ferromagnetism. %
We show that,
when the second top gate is changed to make use of the insulating or spin insulating states,
huge GMR at high temperature ($\sim 10^5\%$ at 68 and 100K) can be achieved for EuO and YIG.
These results imply a magnetic field-free (electrically engineered),
vertical configured, 
and magnetism tunable GMR,
which distinguishes remarkably from the conventional one (see Fig. \ref{mechanism}).
The proposed GMR 
offers not only a viable path to the tantalizing magnetic field-free spintronics,
but also an evidence for the ferromagnetism. 
GMR based on graphene has been widely studied before 
\cite{cheianov2006selective,mccann2006weak,haugen2008spin,zhai2008theory,kim2008prediction,
munoz2009giant,zhang2010spin,lu2011graphene,bai2010very,friedman2010quantum,liao2012large,
mendes2015spin,gopinadhan2015extremely,kisslinger2015linear,zhai2016atomistic,wu2017large,el2017graphene};
however, the value is usually small and a magnetic field is indispensable.

\begin{figure}[tbh]
  \centering
  \includegraphics[width=\linewidth]{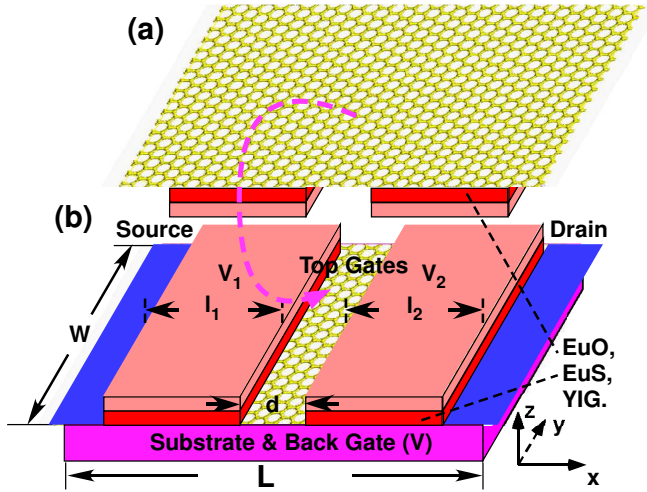}\\
  \caption{The proximity- and gate-induced spin valve.
  (a) Graphene on top of two magnetic insulators contacting with two top gates.
  (b) The spin valve formed by the above structure on a substrate contacting with a back gate.
  }\label{device}
\end{figure}

Figure \ref{device} shows the proximity- and gate-induced spin valve.
Two EuO(111), EuS(111), or YIG(111) substrates of lengths $l_1$ and $l_2$ and a distance $d$ are grown
on top of two `top' gates ($V_1$ and $V_2$).
On the substrates an $L\times W$ \cite{note1} graphene film
is deposited \cite{swartz2012integration,wei2016strong,wang2015proximity}.
The whole structure is then turned over and transferred to
a substrate 
contacting with a back gate ($V$).
The graphene is further contacted with source and drain electrodes ($U$).
As shown by first principle calculations \cite{hallal2017tailoring},
all the ferromagnets are heavily electron doped,
which limits the spintronic application by a low polarization
(about $24\%$ for EuO \cite{yang2013proximity}).
Hole doping by magnetic insulator such as CFO was suggested to overcome this shortcoming \cite{hallal2017tailoring}.
Here we propose a different way, i.e., by applying top gates.
Through the strong electric field effect \cite{novoselov2004electric,kim2012direct},
the Dirac points of the ferromagnets can be tuned to coincide
with the pristine graphene's.

\begin{figure}[tbh]
  \centering
  \includegraphics[width=\linewidth]{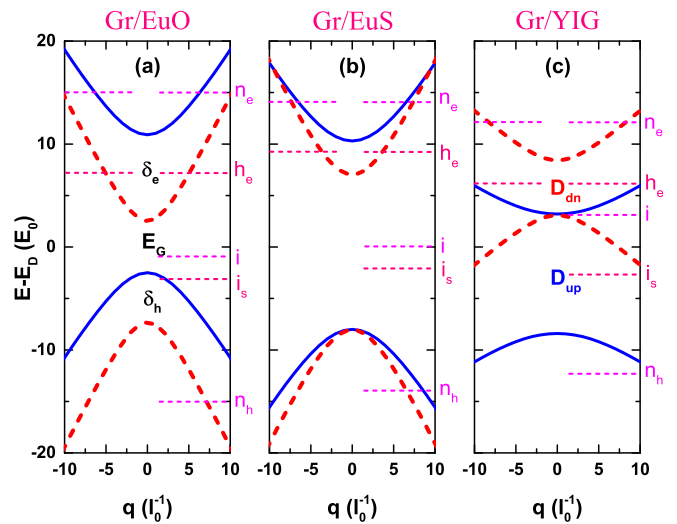}\\
  \caption{Energy dispersions around the Dirac points
  for (a) Gr/EuO, (b) Gr/EuS, and (c) Gr/YIG;
  blue solid for spin up and red dashed for spin down.
  In (a) the Dirac gap ($E_G$) and exchange splittings ($\delta_{e,h}$) are labeled,
  and in (c) the spin Dirac gaps ($D_{up,dn}$) are labeled for comparison.
  The electron and hole normal ferromagnetisms (lines $n_{e,h}$), electron half-metallic ferromagnetism (lines $h_e$),
  insulating (lines $i$), and spin insulating (lines $i_s$) are labeled.
}\label{dispersions}
\end{figure}

The Eu-$4f$ (Fe-$3d$) states in the EuO and EuS (YIG) substrates are polarized.
They overlap with the C-$p_z$ state in graphene and induce the ferromagnetism \cite{yang2013proximity,hallal2017tailoring}.
The predicted energy dispersions of graphene 
on a six-bilayer EuO, EuS, and six-trilayer YIG substrates at the optimized distances
\cite{yang2013proximity} are plotted in Fig. \ref{dispersions}(a)-(c).
Parabolic and spin resolved dispersions are clearly seen,
from which normal and half-metallic ferromagnetisms for electron and hole (see lines labeled by $n_{e,h}$ and $h_e$),
and insulating and spin insulating states ($i$ and $i_s$) can be defined.
For EuS the half-metallic ferromagnetism for hole is absent,
while for YIG the insulating window is rather narrow.
We have proposed to cancel the heavy electron dopings by top gates;
here we propose to make use of the ferromagnetisms by the back gate.
In the left magnet, both normal and half-metallic ferromagnetisms can be utilized
by lifting the Fermi energy into corresponding windows.
(Note this is rather hard for the conventional case.)
We further propose that, by a top gate difference ($\Delta V=V_2-V_1$),
the electron ferromagnetisms in the right magnet can be switched to opposite (hole) ones,
and even insulating and spin insulating states.
These form the basis for the distinguished GMR.

\begin{figure*}
  \centering
  \includegraphics[width=0.8\linewidth]{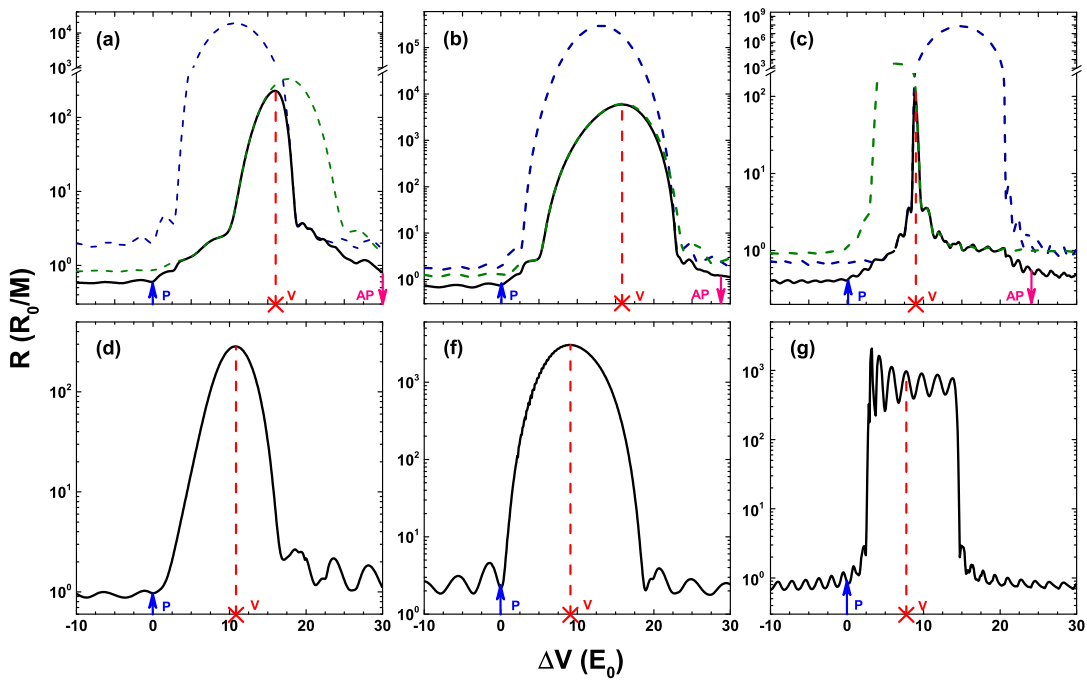}\\
  \caption{MR contributed by normal (a-c) and half-metallic (d-f) ferromagnetisms 
  as a function of voltage difference for EuO, EuS, and YIG.
$V_1=E_D$ and $E_F (V)$ is listed in Table \ref{GMR}.
The P, AP, and V configurations are labeled, which are also
  shown in Fig. \ref{dispersions}.
  In (a-c), according to $R^{-1}=R_\uparrow^{-1}+R_\downarrow^{-1}$, the maximal MR happens between the spin Dirac points
with a value determined by the smaller spin resistance.
 }\label{gates}
\end{figure*}

Heavy electron doping's ($E_{D}$),
band gaps opening at the Dirac points ($E_G$),
and exchange splittings ($\delta_e$ and $\delta_h$)
as labeled in Fig. \ref{dispersions}(a) are observed as the ferromagnetisms.
Accordingly, dispersions around the Dirac points were described
by effective Hamiltonians in a sublattice-spin direct produce space \cite{zollner2016theory,su2017effect,hallal2017tailoring}.
Note, in this space wave function should be solved as a four-components one
even for a single valley \cite{song2013ballistic}.
The ferromagnetisms can also be defined by spin up and spin down ($s=\pm1$)
Dirac cone dopings ($D_s$), Dirac gaps ($\Delta_s$), and Fermi velocities ($v_s$)
as labeled in Fig. \ref{dispersions}(c).
These parameters relate with the above ones by
$D_{\uparrow,\downarrow}=E_D\pm\delta_{e,h}/2$ and 
$\Delta_{\uparrow,\downarrow}=|\delta_{e,h}|+E_G$; 
$v_{\uparrow\downarrow}$
can be fitted from the original data
(see Table \ref{parameter} for parameters).
In this view,
the three magnets can be described by a uniform effective Hamiltonian
in a simple \emph{sublattice space},
\begin{equation}\label{Hamiltonian}
\mathcal{H}_{\bm{k},s,\xi}=\mathcal{I} (E_{Ds} +V_t) + \sigma_z \xi \Delta_s + \bm\sigma \cdot \hbar v_s \bm{k},
\end{equation}
where $\bm{k}=(k_x, k_y)$ is the momentum operator,
$\bm{\sigma}=(\sigma_{x},\sigma _{y})$ is the pseudospin Pauli matrices,
$V_i (i=1,2)$ is the top gate voltages,
$\mathcal{I}$ is the identify matrix,
and $\xi =\pm 1$ for valley $K$ and $K^\prime$.
For pristine and contacted graphene
$\mathcal{H}_{\bm{k},\xi}=\mathcal{I} U + \bm\sigma \cdot \hbar v_F \bm{k}$.

For brevity, we express all quantities in dimensionless form by means of a
characteristic energy $E_0=10$ meV and corresponding length unit
$l_0=\hbar v_F/E_0=56.55$ nm.
The right- and left-going envelope functions ($\Phi_j$) in
the contacted, ferromagnetic, and pristine graphene ($j=c,m,p$)
can be exactly resolved by decoupling
$\mathcal{H}_j\Phi_j=E_j\Phi_j$.
The result reads
$\Phi_j^\pm=[e^{\pm ik_jx},e^{\pm ik_jx}(\pm k_j+iq_j)/E_j]^T e^{iq_jy}/\sqrt{2}$,
where $E_{p(c)}=E(-U)$,
$E_m=(E-D_s-V+\Delta_s)/v_s$,
$q_{p,c,m}=E_e\sin\alpha$, 
$k_{p,c}=\textmd{sign}(E_{p,c})\sqrt{E_{p,c}^{2}-q_{p,c}^{2}}$, and
$k_m=\textmd{sign}(E_m)\sqrt{E_m E_m^\prime-q_m^{2}}$ with $E_m^\prime=(E-D_s-V-\Delta_s)/v_s$.
From the continuity of envelope functions at the boundaries,
transfer matrix $M$ can be constructed \cite{song2012giant,song2013generation} 
and spin-resolved transmission coefficients
can be obtained as $t_s=M[[2,2]]^{-1}$ \cite{born1980principles}.
For $T<$100K, the $e$-$e$ and $e$-$ph$ inelastic scatterings can be ignored \cite{morozov2008giant,chen2008intrinsic}
and the ballistic
spin-resolved conductance can be given
by the Landauer-B\"{u}ttiker formula \cite{buttiker1985generalized}
\begin{equation}\label{T-current}
G_s(T)=G_{0}\int dE\frac{-df}{dE}\int_{-|E_{F}|}^{|E_{F}|}|t_s|^2(E,q)\frac{dq}{2\pi /W},
\end{equation}%
where
$f(E,T)=[1+e^{(E-E_{F})/T}]^{-1}$ is the Fermi-Dirac distribution
function, and $G_{0}=2e^{2}/h$ is the quantum conductance
(2 accounts for the valley degeneracy).
The zero-temperature conductance can be rewritten as
$G_s(0)=MG_{0}\int_{-\pi /2}^{\pi /2}T_s(E_{F},\alpha )\cos \alpha
d\alpha,$
where
$M=(|E_{F}|/E_{0})(W/2\pi l_{0})\equiv M_EM_W$
is half of the number of the transverse modes.
The magnetoresistance (MR) is given by $R=(G_\uparrow+G_\downarrow)^{-1}$
(in unit of $R_0=G_0^{-1}$),
and the GMR is defined by the ratio between the V and P configurations through
$(R_V-R_P)/R_P\times 100\%$.

\begin{figure*}[tbh]
  \centering
  \includegraphics[width=0.9\linewidth]{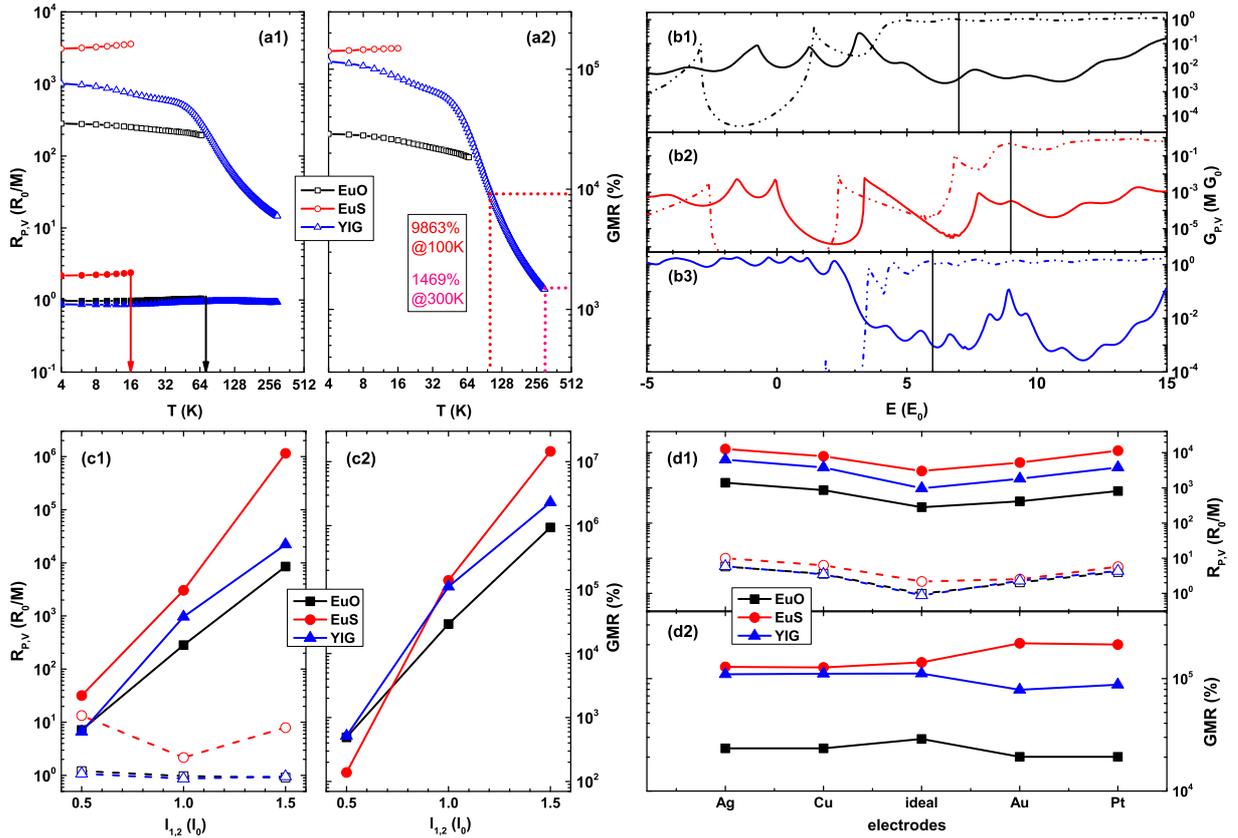}\\
    \caption{MR for the P and V configurations and GMR as a function of (a) temperature,
    (c) magnet length, and (d) electrode doping for EuO, EuS, and YIG.
    (b) Zero temperature MRs as a function of energy for the three magnets and two configurations.
    }\label{temperature}
\end{figure*}

\begin{table}[htbp]
\caption{The spin Dirac doping's, Dirac gaps (both in unit of $E_0$=10meV),
Fermi velocities (in unit of $v_F$), and Curie temperatures \cite{hallal2017tailoring}
for dispersions in Fig. \ref{dispersions}.
}\label{parameter}
\centering
\begin{tabular}{lcccccccc}
\hline
\toprule
magnets & $E_{D\uparrow}$ & $\Delta_\uparrow$ & $v_\uparrow$ & $E_{D\downarrow}$ & $\Delta_\downarrow$ & $v_\downarrow$ & $T_c$ \\
\hline
\midrule
Gr/EuO & -132.8 & 13.4 & 1.337 & -139.4 & 9.8 & 1.628 & 69K \\ 
Gr/EuS & -128.85 & 18.3 & 1.40 & -130.5 & 15 & 1.60 & 16.5K \\ 
Gr/YIG & -80.6 & 11.6 & 0.63 & -83.75 & 5.3 & 0.70 & 550K \\ 
\hline
\bottomrule
\end{tabular}
\end{table}
We first consider GMR utilizing the normal ferromagnetism.
The MR and its spin components ($R_{\uparrow,\downarrow}=G_{\uparrow,\downarrow}^{-1}$) as
a function of the top gate difference are plotted in Fig. \ref{gates}(a)-(c).
As can be seen,
the MR is rather low ($\sim R_0/M$) when a same top gate ($\Delta V=0$)
is applied on the right magnet (see the arrow labeled by P). 
This is the P configuration, for which both spins
transport through the spin valve 
quasi-ballistically.
When a gate difference is applied, the MR first increases and then decreases.
Surprisingly, the maximal MR dose not happens at the AP configuration as the conventional case
(see the arrow labeled by AP).
Instead, it arises when the insulating state in the second magnet
aligns to the Fermi energy (see the arrow labeled by V).
This configuration falls in between the P and AP ones
and can be defined as a \emph{vertical configuration}.
In the right magnets both spins are blocked for the V configuration,
while transport quasi-ballistically again for the AP configuration.
Comparing the MRs at these two configurations, we obtain a rather huge
GMR with typical value of $10^4\%\sim10^5\%$ (see Table \ref{GMR} for details).
It is also seen that, the MR profile for YIG is rather sharp due to
the rather narrow insulating window.

We then consider GMR utilizing the half-metallic ferromagnetism, for which
only one spin contributes the MR.
The MR$-\Delta V$ profiles are plotted in Fig. \ref{gates} (d)-(g).
It is not surprised that, rather low MR for
the P configuration and rather high MR for the V configuration
are observed again.
Comparing them, a rather huge GMR
with typical value of $10^4\%\sim10^5\%$ are obtained (see table \ref{GMR}).
Interestingly, the high resistance for YIG shows rich resonant peaks.
This is due to a much smaller Dirac gap for spin down, which supports
much smaller resistance even for a blocked-blocked transport.
It seems that, an AP configuration is responsible for the EuO and YIG cases.
However, 
since spin up or spin down is always blocked,
only the blocking of the initially transparent spin (spin down or up) is important.
This is confirmed by the EuS case, where
GMR happens in absence of opposite (hole) half-metallic ferromagnetism.

In conventional spin valves made of no matter normal \cite{baibich1988giant} or half-metallic  \cite{hwang1997enhanced}
ferromagnets,
P and AP configurations are respectively responsible for the low and high MRs;
they are switched by a magnetic field (see the left column in Fig. \ref{mechanism}).
The mechanism for the proximity- and gate-induced graphene spin valve is totally different.
The V configuration is responsible for the high resistance states
and it is switched from the P configuration by an electric field.
Moreover, normal and half-metallic ferromagnetisms can also be
switched by the back gate, which is usually forbidden for the conventional cases
(see the right column in Fig. \ref{mechanism})

\begin{table}[htbp]
\caption{GMR and low and high MRs, corresponding
ferromagnetisms and voltage difference summarized from Fig. \ref{gates}.
}\label{GMR}
\centering
\begin{tabular}{clll}
\hline
\toprule
$GMR$ & $R_{AP}/R_P$ & ferromagnetisms ($E_F$) &  $\Delta V_V$\\
\hline
\midrule
$3.8\times 10^4\%$ & 228.4/0.603 & EuO normal (15) & 16\\
$7.9\times 10^5\%$ & 5935/0.748 & EuS normal (14) & 15.9\\
$2.6\times 10^4\%$ & 108/0.411 & YIG normal (12) & 8.9\\
$2.9\times 10^4\%$ & 278.5/0.971 & EuO half-metallic (7) & 10.45\\
$1.4\times 10^5\%$ & 3017/2.18 & EuS half-metallic (9) & 8.95\\
$1.1\times 10^5\%$ & 968/0.845 & YIG half-metallic (6) & 7.75\\
\hline
\bottomrule
\end{tabular}
\end{table}

Application at high temperature is crucial.
Fig. \ref{temperature}(a) shows
$R_P$, 
$R_V$, 
and GMR as a function of temperature.
The temperature range is limited to
$\min(100K, T_c)$,
for which the ferromagnetisms hold and the inelastic scattering
can be ignored. 
It is seen that,
for all ferromagnets $R_P$ increase slightly as the temperature,
while $R_V$ shows a complicated dependence.
It decreases with temperature for EuO and YIG and increases for EuS.
Accordingly, the GMR for EuS/EuO/YIG follows an increasing/decreasing/decreasing behavior,
with a value of $1.5\times 10^5\%$/$1.8\times 10^4\%$/$9.8\times 10^3\%$
observed at 16K/68K/100K.
For YIG the Curie temperature is higher than room temperature.
A large GMR of $1470\%$ is evaluated at 300K by ignoring the inelastic scattering \cite{note3}.
These values imply promising high temperature or even room temperature GMR
for EuO and YIG.

Due to Eq. (\ref{T-current}), spin current at a finite temperature $T$ is determined by the
zero-temperature ones in an energy range $\sim(E_{F}-5T,E_{F}+5T)$.
In Fig. \ref{temperature}(b), we plot the latter
for the three magnets and two configurations.
It can be seen that,
the $P$ ($V$) conductance for all magnets (EuS)
reaches almost the maximum around the Fermi energy.
As a result,
the higher the temperature, 
the smaller (bigger) the finite temperature spin current (MR).
The cases become opposite for the V configuration of EuO and YIG,
because the spin current reaches almost the local minimum at the Fermi energy.
The different energy bands
are responsible for these different temperature dependences.

Fig. \ref{temperature}(c) shows the dependence of $R_P$,
$R_V$, and GMR on the magnet length.
It is found that, $R_P$ changes slightly with the magnet length
while $R_V$ increases exponentially.
When the length increase from 1 to 1.5 (still within the ballistic regime),
the GMR increases 
to $0.94\times 10^6\%$, $1.4\times 10^7\%$, $2.3\times 10^6\%$ for EuO, EuS, and YIG,
respectively.
This is as large as 
the extraordinary magnetoresistance in semiconductor-metal hybrid systems \cite{solin2000enhanced,hewett2010geometrically}.
For the V configuration, electrons transport evanescently in the second magnet.
Due to an imaginary wave vector $k_m$ in the term $e^{ik_mx}$,
this leads a behavior of $|t|^2\sim e^{-2l}$ and $R\sim e^{2l}$.
Due to this behavior, the negative temperature dependence 
(see Fig. \ref{temperature} (a))
for YIG and EuO can be counteracted, and a huge room temperature GMR is expectable for YIG.
Asymmetric spin valves with short left and long right magnets are suggested to enhance the GMR
within the ballistic regime.

In all the above calculations, the effect of electrodes is ignored.
Such an ideal contact can be achieved by specific metals with special
distance to graphene (e.g., Au/Cu/Ag at 3.2/3.4/3.7${\AA}$) \cite{giovannetti2008doping}.
Fig. \ref{temperature}(d) shows the calculated results for several
familiar metal electrodes at their equilibrium distances.
For Ag, Cu, Au, and Pt, $U/E_0=-32,-17,19,32$ respectively \cite{giovannetti2008doping},
which are rather smaller than the proximity induced doping's.
It is observed that, no matter for which magnets and for which configurations,
the MR increases as the electrode becomes non ideal;
the heavier the contact doping, the larger the MR increases.
This is because the symmetric pristine-ferromagnetism (insulating)-pristine structure
for the left (right) magnets becomes asymmetric doped-ferromagnetism-pristine or pristine-insulating-doped ones.
However, the GMR can show different behaviors.
It increases for EuS with positive doping (Au and Pt) and
for YIG with negative doping (Cu and Ag),
while decreases for EuO with any doping and
for EuS (YIG) with negative (positive) doping.
These results suggest that, the proposed GMR is rather robust
to familiar metallic contacts, and can be even enhanced for YIG and EuS
with proper contacts.

In summary, we have proposed a distinguished GMR that is
magnetic field-free, vertically configured, and magnetism-tunable.
The proximity effect 
and electric field effect 
in a novel graphene spin valve, 
both stemming from the 2D nature of graphene,
play a central role.
Outstanding performances such as
huge values at high temperature ($\sim 10^5\%$ at 68K and 100K),
exponential enhancement by magnet length,
and robustness to familiar electrodes
have been demonstrated.
These results may offer a viable path to a magnetic field-free spintronics
as well as an evidence for the magnetisms.
The uniform Hamiltonian constructed in the sublattice space
can be applied to investigate spin transport 
in related nanostructures.
Since the spin Dirac gaps increases as fewer layer \cite{hallal2017tailoring},
larger GMR is predicted in graphene grown on thinner substrates.
We encourage experimental researches on the proposed GMR and mechanisms.

This work was supported by the National Natural Science Foundation of China (NSFC) under Grant No. 11404300,
the Science Challenge Project (SCP) under Grant No. TZ2016003-1,
and the S$\&$T Innovation Fund of IEE, CAEP under Grant No. S20140807.


\begin{thebibliography}{47}%
\makeatletter
\providecommand \@ifxundefined [1]{%
 \@ifx{#1\undefined}
}%
\providecommand \@ifnum [1]{%
 \ifnum #1\expandafter \@firstoftwo
 \else \expandafter \@secondoftwo
 \fi
}%
\providecommand \@ifx [1]{%
 \ifx #1\expandafter \@firstoftwo
 \else \expandafter \@secondoftwo
 \fi
}%
\providecommand \natexlab [1]{#1}%
\providecommand \enquote  [1]{``#1''}%
\providecommand \bibnamefont  [1]{#1}%
\providecommand \bibfnamefont [1]{#1}%
\providecommand \citenamefont [1]{#1}%
\providecommand \href@noop [0]{\@secondoftwo}%
\providecommand \href [0]{\begingroup \@sanitize@url \@href}%
\providecommand \@href[1]{\@@startlink{#1}\@@href}%
\providecommand \@@href[1]{\endgroup#1\@@endlink}%
\providecommand \@sanitize@url [0]{\catcode `\\12\catcode `\$12\catcode
  `\&12\catcode `\#12\catcode `\^12\catcode `\_12\catcode `\%12\relax}%
\providecommand \@@startlink[1]{}%
\providecommand \@@endlink[0]{}%
\providecommand \url  [0]{\begingroup\@sanitize@url \@url }%
\providecommand \@url [1]{\endgroup\@href {#1}{\urlprefix }}%
\providecommand \urlprefix  [0]{URL }%
\providecommand \Eprint [0]{\href }%
\providecommand \doibase [0]{http://dx.doi.org/}%
\providecommand \selectlanguage [0]{\@gobble}%
\providecommand \bibinfo  [0]{\@secondoftwo}%
\providecommand \bibfield  [0]{\@secondoftwo}%
\providecommand \translation [1]{[#1]}%
\providecommand \BibitemOpen [0]{}%
\providecommand \bibitemStop [0]{}%
\providecommand \bibitemNoStop [0]{.\EOS\space}%
\providecommand \EOS [0]{\spacefactor3000\relax}%
\providecommand \BibitemShut  [1]{\csname bibitem#1\endcsname}%
\let\auto@bib@innerbib\@empty
\bibitem [{\citenamefont {Han}\ \emph {et~al.}(2014)\citenamefont {Han},
  \citenamefont {Kawakami}, \citenamefont {Gmitra},\ and\ \citenamefont
  {Fabian}}]{han2014graphene}%
  \BibitemOpen
  \bibfield  {author} {\bibinfo {author} {\bibfnamefont {W.}~\bibnamefont
  {Han}}, \bibinfo {author} {\bibfnamefont {R.~K.}\ \bibnamefont {Kawakami}},
  \bibinfo {author} {\bibfnamefont {M.}~\bibnamefont {Gmitra}}, \ and\ \bibinfo
  {author} {\bibfnamefont {J.}~\bibnamefont {Fabian}},\ }\href@noop {}
  {\bibfield  {journal} {\bibinfo  {journal} {Nature Nanotechnology}\ }\textbf
  {\bibinfo {volume} {9}},\ \bibinfo {pages} {794} (\bibinfo {year}
  {2014})}\BibitemShut {NoStop}%
\bibitem [{\citenamefont {Feng}\ \emph {et~al.}(2017)\citenamefont {Feng},
  \citenamefont {Shen}, \citenamefont {Yang}, \citenamefont {Wang},
  \citenamefont {Zeng}, \citenamefont {Wu}, \citenamefont {Chintalapati},\ and\
  \citenamefont {Chang}}]{feng2017prospects}%
  \BibitemOpen
  \bibfield  {author} {\bibinfo {author} {\bibfnamefont {Y.~P.}\ \bibnamefont
  {Feng}}, \bibinfo {author} {\bibfnamefont {L.}~\bibnamefont {Shen}}, \bibinfo
  {author} {\bibfnamefont {M.}~\bibnamefont {Yang}}, \bibinfo {author}
  {\bibfnamefont {A.}~\bibnamefont {Wang}}, \bibinfo {author} {\bibfnamefont
  {M.}~\bibnamefont {Zeng}}, \bibinfo {author} {\bibfnamefont {Q.}~\bibnamefont
  {Wu}}, \bibinfo {author} {\bibfnamefont {S.}~\bibnamefont {Chintalapati}}, \
  and\ \bibinfo {author} {\bibfnamefont {C.-R.}\ \bibnamefont {Chang}},\
  }\href@noop {} {\bibfield  {journal} {\bibinfo  {journal} {Wiley
  Interdisciplinary Reviews: Computational Molecular Science}\ } (\bibinfo
  {year} {2017})}\BibitemShut {NoStop}%
\bibitem [{\citenamefont {Haugen}\ \emph {et~al.}(2008)\citenamefont {Haugen},
  \citenamefont {Huertas-Hernando},\ and\ \citenamefont
  {Brataas}}]{haugen2008spin}%
  \BibitemOpen
  \bibfield  {author} {\bibinfo {author} {\bibfnamefont {H.}~\bibnamefont
  {Haugen}}, \bibinfo {author} {\bibfnamefont {D.}~\bibnamefont
  {Huertas-Hernando}}, \ and\ \bibinfo {author} {\bibfnamefont
  {A.}~\bibnamefont {Brataas}},\ }\href@noop {} {\bibfield  {journal} {\bibinfo
   {journal} {Physical Review B}\ }\textbf {\bibinfo {volume} {77}},\ \bibinfo
  {pages} {115406} (\bibinfo {year} {2008})}\BibitemShut {NoStop}%
\bibitem [{\citenamefont {Yang}\ \emph {et~al.}(2013)\citenamefont {Yang},
  \citenamefont {Hallal}, \citenamefont {Terrade}, \citenamefont {Waintal},
  \citenamefont {Roche},\ and\ \citenamefont {Chshiev}}]{yang2013proximity}%
  \BibitemOpen
  \bibfield  {author} {\bibinfo {author} {\bibfnamefont {H.-X.}\ \bibnamefont
  {Yang}}, \bibinfo {author} {\bibfnamefont {A.}~\bibnamefont {Hallal}},
  \bibinfo {author} {\bibfnamefont {D.}~\bibnamefont {Terrade}}, \bibinfo
  {author} {\bibfnamefont {X.}~\bibnamefont {Waintal}}, \bibinfo {author}
  {\bibfnamefont {S.}~\bibnamefont {Roche}}, \ and\ \bibinfo {author}
  {\bibfnamefont {M.}~\bibnamefont {Chshiev}},\ }\href@noop {} {\bibfield
  {journal} {\bibinfo  {journal} {Physical Review Letters}\ }\textbf {\bibinfo
  {volume} {110}},\ \bibinfo {pages} {046603} (\bibinfo {year}
  {2013})}\BibitemShut {NoStop}%
\bibitem [{\citenamefont {Hallal}\ \emph {et~al.}(2017)\citenamefont {Hallal},
  \citenamefont {Ibrahim}, \citenamefont {Yang}, \citenamefont {Roche},\ and\
  \citenamefont {Chshiev}}]{hallal2017tailoring}%
  \BibitemOpen
  \bibfield  {author} {\bibinfo {author} {\bibfnamefont {A.}~\bibnamefont
  {Hallal}}, \bibinfo {author} {\bibfnamefont {F.}~\bibnamefont {Ibrahim}},
  \bibinfo {author} {\bibfnamefont {H.}~\bibnamefont {Yang}}, \bibinfo {author}
  {\bibfnamefont {S.}~\bibnamefont {Roche}}, \ and\ \bibinfo {author}
  {\bibfnamefont {M.}~\bibnamefont {Chshiev}},\ }\href@noop {} {\bibfield
  {journal} {\bibinfo  {journal} {2D Materials}\ } (\bibinfo {year}
  {2017})}\BibitemShut {NoStop}%
\bibitem [{\citenamefont {Song}\ and\ \citenamefont
  {Dai}(2015)}]{song2015spin}%
  \BibitemOpen
  \bibfield  {author} {\bibinfo {author} {\bibfnamefont {Y.}~\bibnamefont
  {Song}}\ and\ \bibinfo {author} {\bibfnamefont {G.}~\bibnamefont {Dai}},\
  }\href@noop {} {\bibfield  {journal} {\bibinfo  {journal} {Applied Physics
  Letters}\ }\textbf {\bibinfo {volume} {106}},\ \bibinfo {pages} {223104}
  (\bibinfo {year} {2015})}\BibitemShut {NoStop}%
\bibitem [{\citenamefont {Ang}\ \emph {et~al.}(2016)\citenamefont {Ang},
  \citenamefont {Ang}, \citenamefont {Zhang},\ and\ \citenamefont
  {Ma}}]{ang2016nonlocal}%
  \BibitemOpen
  \bibfield  {author} {\bibinfo {author} {\bibfnamefont {Y.~S.}\ \bibnamefont
  {Ang}}, \bibinfo {author} {\bibfnamefont {L.}~\bibnamefont {Ang}}, \bibinfo
  {author} {\bibfnamefont {C.}~\bibnamefont {Zhang}}, \ and\ \bibinfo {author}
  {\bibfnamefont {Z.}~\bibnamefont {Ma}},\ }\href@noop {} {\bibfield  {journal}
  {\bibinfo  {journal} {Physical Review B}\ }\textbf {\bibinfo {volume} {93}},\
  \bibinfo {pages} {041422} (\bibinfo {year} {2016})}\BibitemShut {NoStop}%
\bibitem [{\citenamefont {Su}\ \emph {et~al.}(2017)\citenamefont {Su},
  \citenamefont {Barlas}, \citenamefont {Li}, \citenamefont {Shi},\ and\
  \citenamefont {Lake}}]{su2017effect}%
  \BibitemOpen
  \bibfield  {author} {\bibinfo {author} {\bibfnamefont {S.}~\bibnamefont
  {Su}}, \bibinfo {author} {\bibfnamefont {Y.}~\bibnamefont {Barlas}}, \bibinfo
  {author} {\bibfnamefont {J.}~\bibnamefont {Li}}, \bibinfo {author}
  {\bibfnamefont {J.}~\bibnamefont {Shi}}, \ and\ \bibinfo {author}
  {\bibfnamefont {R.~K.}\ \bibnamefont {Lake}},\ }\href@noop {} {\bibfield
  {journal} {\bibinfo  {journal} {Physical Review B}\ }\textbf {\bibinfo
  {volume} {95}},\ \bibinfo {pages} {075418} (\bibinfo {year}
  {2017})}\BibitemShut {NoStop}%
\bibitem [{\citenamefont {Swartz}\ \emph {et~al.}(2012)\citenamefont {Swartz},
  \citenamefont {Odenthal}, \citenamefont {Hao}, \citenamefont {Ruoff},\ and\
  \citenamefont {Kawakami}}]{swartz2012integration}%
  \BibitemOpen
  \bibfield  {author} {\bibinfo {author} {\bibfnamefont {A.~G.}\ \bibnamefont
  {Swartz}}, \bibinfo {author} {\bibfnamefont {P.~M.}\ \bibnamefont
  {Odenthal}}, \bibinfo {author} {\bibfnamefont {Y.}~\bibnamefont {Hao}},
  \bibinfo {author} {\bibfnamefont {R.~S.}\ \bibnamefont {Ruoff}}, \ and\
  \bibinfo {author} {\bibfnamefont {R.~K.}\ \bibnamefont {Kawakami}},\
  }\href@noop {} {\bibfield  {journal} {\bibinfo  {journal} {ACS Nano}\
  }\textbf {\bibinfo {volume} {6}},\ \bibinfo {pages} {10063} (\bibinfo {year}
  {2012})}\BibitemShut {NoStop}%
\bibitem [{\citenamefont {Swartz}\ \emph {et~al.}(2013)\citenamefont {Swartz},
  \citenamefont {McCreary}, \citenamefont {Han}, \citenamefont {Wong},
  \citenamefont {Odenthal}, \citenamefont {Wen}, \citenamefont {Chen},
  \citenamefont {Kawakami}, \citenamefont {Hao}, \citenamefont {Ruoff} \emph
  {et~al.}}]{swartz2013integrating}%
  \BibitemOpen
  \bibfield  {author} {\bibinfo {author} {\bibfnamefont {A.~G.}\ \bibnamefont
  {Swartz}}, \bibinfo {author} {\bibfnamefont {K.~M.}\ \bibnamefont
  {McCreary}}, \bibinfo {author} {\bibfnamefont {W.}~\bibnamefont {Han}},
  \bibinfo {author} {\bibfnamefont {J.~J.}\ \bibnamefont {Wong}}, \bibinfo
  {author} {\bibfnamefont {P.~M.}\ \bibnamefont {Odenthal}}, \bibinfo {author}
  {\bibfnamefont {H.}~\bibnamefont {Wen}}, \bibinfo {author} {\bibfnamefont
  {J.-R.}\ \bibnamefont {Chen}}, \bibinfo {author} {\bibfnamefont {R.~K.}\
  \bibnamefont {Kawakami}}, \bibinfo {author} {\bibfnamefont {Y.}~\bibnamefont
  {Hao}}, \bibinfo {author} {\bibfnamefont {R.~S.}\ \bibnamefont {Ruoff}},
  \emph {et~al.},\ }\href@noop {} {\bibfield  {journal} {\bibinfo  {journal}
  {Journal of Vacuum Science \& Technology B, Nanotechnology and
  Microelectronics: Materials, Processing, Measurement, and Phenomena}\
  }\textbf {\bibinfo {volume} {31}},\ \bibinfo {pages} {04D105} (\bibinfo
  {year} {2013})}\BibitemShut {NoStop}%
\bibitem [{\citenamefont {Wang}\ \emph {et~al.}(2015)\citenamefont {Wang},
  \citenamefont {Tang}, \citenamefont {Sachs}, \citenamefont {Barlas},\ and\
  \citenamefont {Shi}}]{wang2015proximity}%
  \BibitemOpen
  \bibfield  {author} {\bibinfo {author} {\bibfnamefont {Z.}~\bibnamefont
  {Wang}}, \bibinfo {author} {\bibfnamefont {C.}~\bibnamefont {Tang}}, \bibinfo
  {author} {\bibfnamefont {R.}~\bibnamefont {Sachs}}, \bibinfo {author}
  {\bibfnamefont {Y.}~\bibnamefont {Barlas}}, \ and\ \bibinfo {author}
  {\bibfnamefont {J.}~\bibnamefont {Shi}},\ }\href@noop {} {\bibfield
  {journal} {\bibinfo  {journal} {Physical Review Letters}\ }\textbf {\bibinfo
  {volume} {114}},\ \bibinfo {pages} {016603} (\bibinfo {year}
  {2015})}\BibitemShut {NoStop}%
\bibitem [{\citenamefont {Mendes}\ \emph {et~al.}(2015)\citenamefont {Mendes},
  \citenamefont {Santos}, \citenamefont {Meireles}, \citenamefont {Lacerda},
  \citenamefont {Vilela-Le{\~a}o}, \citenamefont {Machado}, \citenamefont
  {Rodr{\'\i}guez-Su{\'a}rez}, \citenamefont {Azevedo},\ and\ \citenamefont
  {Rezende}}]{mendes2015spin}%
  \BibitemOpen
  \bibfield  {author} {\bibinfo {author} {\bibfnamefont {J.}~\bibnamefont
  {Mendes}}, \bibinfo {author} {\bibfnamefont {O.~A.}\ \bibnamefont {Santos}},
  \bibinfo {author} {\bibfnamefont {L.}~\bibnamefont {Meireles}}, \bibinfo
  {author} {\bibfnamefont {R.}~\bibnamefont {Lacerda}}, \bibinfo {author}
  {\bibfnamefont {L.}~\bibnamefont {Vilela-Le{\~a}o}}, \bibinfo {author}
  {\bibfnamefont {F.}~\bibnamefont {Machado}}, \bibinfo {author} {\bibfnamefont
  {R.}~\bibnamefont {Rodr{\'\i}guez-Su{\'a}rez}}, \bibinfo {author}
  {\bibfnamefont {A.}~\bibnamefont {Azevedo}}, \ and\ \bibinfo {author}
  {\bibfnamefont {S.}~\bibnamefont {Rezende}},\ }\href@noop {} {\bibfield
  {journal} {\bibinfo  {journal} {Physical Review Letters}\ }\textbf {\bibinfo
  {volume} {115}},\ \bibinfo {pages} {226601} (\bibinfo {year}
  {2015})}\BibitemShut {NoStop}%
\bibitem [{\citenamefont {Leutenantsmeyer}\ \emph {et~al.}(2016)\citenamefont
  {Leutenantsmeyer}, \citenamefont {Kaverzin}, \citenamefont {Wojtaszek},\ and\
  \citenamefont {van Wees}}]{leutenantsmeyer2016proximity}%
  \BibitemOpen
  \bibfield  {author} {\bibinfo {author} {\bibfnamefont {J.~C.}\ \bibnamefont
  {Leutenantsmeyer}}, \bibinfo {author} {\bibfnamefont {A.~A.}\ \bibnamefont
  {Kaverzin}}, \bibinfo {author} {\bibfnamefont {M.}~\bibnamefont {Wojtaszek}},
  \ and\ \bibinfo {author} {\bibfnamefont {B.~J.}\ \bibnamefont {van Wees}},\
  }\href@noop {} {\bibfield  {journal} {\bibinfo  {journal} {2D Materials}\
  }\textbf {\bibinfo {volume} {4}},\ \bibinfo {pages} {014001} (\bibinfo {year}
  {2016})}\BibitemShut {NoStop}%
\bibitem [{\citenamefont {Evelt}\ \emph {et~al.}(2017)\citenamefont {Evelt},
  \citenamefont {Ochoa}, \citenamefont {Dzyapko}, \citenamefont {Demidov},
  \citenamefont {Yurgens}, \citenamefont {Sun}, \citenamefont {Tserkovnyak},
  \citenamefont {Bessonov}, \citenamefont {Rinkevich},\ and\ \citenamefont
  {Demokritov}}]{evelt2017chiral}%
  \BibitemOpen
  \bibfield  {author} {\bibinfo {author} {\bibfnamefont {M.}~\bibnamefont
  {Evelt}}, \bibinfo {author} {\bibfnamefont {H.}~\bibnamefont {Ochoa}},
  \bibinfo {author} {\bibfnamefont {O.}~\bibnamefont {Dzyapko}}, \bibinfo
  {author} {\bibfnamefont {V.~E.}\ \bibnamefont {Demidov}}, \bibinfo {author}
  {\bibfnamefont {A.}~\bibnamefont {Yurgens}}, \bibinfo {author} {\bibfnamefont
  {J.}~\bibnamefont {Sun}}, \bibinfo {author} {\bibfnamefont {Y.}~\bibnamefont
  {Tserkovnyak}}, \bibinfo {author} {\bibfnamefont {V.}~\bibnamefont
  {Bessonov}}, \bibinfo {author} {\bibfnamefont {A.~B.}\ \bibnamefont
  {Rinkevich}}, \ and\ \bibinfo {author} {\bibfnamefont {S.~O.}\ \bibnamefont
  {Demokritov}},\ }\href@noop {} {\bibfield  {journal} {\bibinfo  {journal}
  {Physical Review B}\ }\textbf {\bibinfo {volume} {95}},\ \bibinfo {pages}
  {024408} (\bibinfo {year} {2017})}\BibitemShut {NoStop}%
\bibitem [{\citenamefont {Wei}\ \emph {et~al.}(2016)\citenamefont {Wei},
  \citenamefont {Lee}, \citenamefont {Lemaitre}, \citenamefont {Pinel},
  \citenamefont {Cutaia}, \citenamefont {Cha}, \citenamefont {Katmis},
  \citenamefont {Zhu}, \citenamefont {Heiman}, \citenamefont {Hone} \emph
  {et~al.}}]{wei2016strong}%
  \BibitemOpen
  \bibfield  {author} {\bibinfo {author} {\bibfnamefont {P.}~\bibnamefont
  {Wei}}, \bibinfo {author} {\bibfnamefont {S.}~\bibnamefont {Lee}}, \bibinfo
  {author} {\bibfnamefont {F.}~\bibnamefont {Lemaitre}}, \bibinfo {author}
  {\bibfnamefont {L.}~\bibnamefont {Pinel}}, \bibinfo {author} {\bibfnamefont
  {D.}~\bibnamefont {Cutaia}}, \bibinfo {author} {\bibfnamefont
  {W.}~\bibnamefont {Cha}}, \bibinfo {author} {\bibfnamefont {F.}~\bibnamefont
  {Katmis}}, \bibinfo {author} {\bibfnamefont {Y.}~\bibnamefont {Zhu}},
  \bibinfo {author} {\bibfnamefont {D.}~\bibnamefont {Heiman}}, \bibinfo
  {author} {\bibfnamefont {J.}~\bibnamefont {Hone}},  \emph {et~al.},\
  }\href@noop {} {\bibfield  {journal} {\bibinfo  {journal} {Nature Materials}\
  } (\bibinfo {year} {2016})}\BibitemShut {NoStop}%
\bibitem [{\citenamefont {Novoselov}\ \emph {et~al.}(2004)\citenamefont
  {Novoselov}, \citenamefont {Geim}, \citenamefont {Morozov}, \citenamefont
  {Jiang}, \citenamefont {Zhang}, \citenamefont {Dubonos}, \citenamefont
  {Grigorieva},\ and\ \citenamefont {Firsov}}]{novoselov2004electric}%
  \BibitemOpen
  \bibfield  {author} {\bibinfo {author} {\bibfnamefont {K.~S.}\ \bibnamefont
  {Novoselov}}, \bibinfo {author} {\bibfnamefont {A.~K.}\ \bibnamefont {Geim}},
  \bibinfo {author} {\bibfnamefont {S.~V.}\ \bibnamefont {Morozov}}, \bibinfo
  {author} {\bibfnamefont {D.}~\bibnamefont {Jiang}}, \bibinfo {author}
  {\bibfnamefont {Y.}~\bibnamefont {Zhang}}, \bibinfo {author} {\bibfnamefont
  {S.~V.}\ \bibnamefont {Dubonos}}, \bibinfo {author} {\bibfnamefont {I.~V.}\
  \bibnamefont {Grigorieva}}, \ and\ \bibinfo {author} {\bibfnamefont {A.~A.}\
  \bibnamefont {Firsov}},\ }\href@noop {} {\bibfield  {journal} {\bibinfo
  {journal} {Science}\ }\textbf {\bibinfo {volume} {306}},\ \bibinfo {pages}
  {666} (\bibinfo {year} {2004})}\BibitemShut {NoStop}%
\bibitem [{\citenamefont {Kim}\ \emph {et~al.}(2012)\citenamefont {Kim},
  \citenamefont {Jo}, \citenamefont {Dillen}, \citenamefont {Ferrer},
  \citenamefont {Fallahazad}, \citenamefont {Yao}, \citenamefont {Banerjee},\
  and\ \citenamefont {Tutuc}}]{kim2012direct}%
  \BibitemOpen
  \bibfield  {author} {\bibinfo {author} {\bibfnamefont {S.}~\bibnamefont
  {Kim}}, \bibinfo {author} {\bibfnamefont {I.}~\bibnamefont {Jo}}, \bibinfo
  {author} {\bibfnamefont {D.}~\bibnamefont {Dillen}}, \bibinfo {author}
  {\bibfnamefont {D.}~\bibnamefont {Ferrer}}, \bibinfo {author} {\bibfnamefont
  {B.}~\bibnamefont {Fallahazad}}, \bibinfo {author} {\bibfnamefont
  {Z.}~\bibnamefont {Yao}}, \bibinfo {author} {\bibfnamefont {S.}~\bibnamefont
  {Banerjee}}, \ and\ \bibinfo {author} {\bibfnamefont {E.}~\bibnamefont
  {Tutuc}},\ }\href@noop {} {\bibfield  {journal} {\bibinfo  {journal}
  {Physical Review Letters}\ }\textbf {\bibinfo {volume} {108}},\ \bibinfo
  {pages} {116404} (\bibinfo {year} {2012})}\BibitemShut {NoStop}%
\bibitem [{\citenamefont {Baibich}\ \emph {et~al.}(1988)\citenamefont
  {Baibich}, \citenamefont {Broto}, \citenamefont {Fert}, \citenamefont
  {Van~Dau}, \citenamefont {Petroff}, \citenamefont {Etienne}, \citenamefont
  {Creuzet}, \citenamefont {Friederich},\ and\ \citenamefont
  {Chazelas}}]{baibich1988giant}%
  \BibitemOpen
  \bibfield  {author} {\bibinfo {author} {\bibfnamefont {M.~N.}\ \bibnamefont
  {Baibich}}, \bibinfo {author} {\bibfnamefont {J.~M.}\ \bibnamefont {Broto}},
  \bibinfo {author} {\bibfnamefont {A.}~\bibnamefont {Fert}}, \bibinfo {author}
  {\bibfnamefont {F.~N.}\ \bibnamefont {Van~Dau}}, \bibinfo {author}
  {\bibfnamefont {F.}~\bibnamefont {Petroff}}, \bibinfo {author} {\bibfnamefont
  {P.}~\bibnamefont {Etienne}}, \bibinfo {author} {\bibfnamefont
  {G.}~\bibnamefont {Creuzet}}, \bibinfo {author} {\bibfnamefont
  {A.}~\bibnamefont {Friederich}}, \ and\ \bibinfo {author} {\bibfnamefont
  {J.}~\bibnamefont {Chazelas}},\ }\href@noop {} {\bibfield  {journal}
  {\bibinfo  {journal} {Physical Review Letters}\ }\textbf {\bibinfo {volume}
  {61}},\ \bibinfo {pages} {2472} (\bibinfo {year} {1988})}\BibitemShut
  {NoStop}%
\bibitem [{\citenamefont {Hwang}\ and\ \citenamefont
  {Cheong}(1997)}]{hwang1997enhanced}%
  \BibitemOpen
  \bibfield  {author} {\bibinfo {author} {\bibfnamefont {H.}~\bibnamefont
  {Hwang}}\ and\ \bibinfo {author} {\bibfnamefont {S.-W.}\ \bibnamefont
  {Cheong}},\ }\href@noop {} {\bibfield  {journal} {\bibinfo  {journal}
  {Science}\ }\textbf {\bibinfo {volume} {278}},\ \bibinfo {pages} {1607}
  (\bibinfo {year} {1997})}\BibitemShut {NoStop}%
\bibitem [{\citenamefont {Cheianov}\ and\ \citenamefont
  {Fal��ko}(2006)}]{cheianov2006selective}%
  \BibitemOpen
  \bibfield  {author} {\bibinfo {author} {\bibfnamefont {V.~V.}\ \bibnamefont
  {Cheianov}}\ and\ \bibinfo {author} {\bibfnamefont {V.~I.}\ \bibnamefont
  {Fal��ko}},\ }\href@noop {} {\bibfield  {journal} {\bibinfo  {journal}
  {Physical Review B}\ }\textbf {\bibinfo {volume} {74}},\ \bibinfo {pages}
  {041403} (\bibinfo {year} {2006})}\BibitemShut {NoStop}%
\bibitem [{\citenamefont {McCann}\ \emph {et~al.}(2006)\citenamefont {McCann},
  \citenamefont {Kechedzhi}, \citenamefont {Fal��ko}, \citenamefont {Suzuura},
  \citenamefont {Ando},\ and\ \citenamefont {Altshuler}}]{mccann2006weak}%
  \BibitemOpen
  \bibfield  {author} {\bibinfo {author} {\bibfnamefont {E.}~\bibnamefont
  {McCann}}, \bibinfo {author} {\bibfnamefont {K.}~\bibnamefont {Kechedzhi}},
  \bibinfo {author} {\bibfnamefont {V.~I.}\ \bibnamefont {Fal��ko}}, \bibinfo
  {author} {\bibfnamefont {H.}~\bibnamefont {Suzuura}}, \bibinfo {author}
  {\bibfnamefont {T.}~\bibnamefont {Ando}}, \ and\ \bibinfo {author}
  {\bibfnamefont {B.}~\bibnamefont {Altshuler}},\ }\href@noop {} {\bibfield
  {journal} {\bibinfo  {journal} {Physical Review Letters}\ }\textbf {\bibinfo
  {volume} {97}},\ \bibinfo {pages} {146805} (\bibinfo {year}
  {2006})}\BibitemShut {NoStop}%
\bibitem [{\citenamefont {Zhai}\ and\ \citenamefont
  {Chang}(2008)}]{zhai2008theory}%
  \BibitemOpen
  \bibfield  {author} {\bibinfo {author} {\bibfnamefont {F.}~\bibnamefont
  {Zhai}}\ and\ \bibinfo {author} {\bibfnamefont {K.}~\bibnamefont {Chang}},\
  }\href@noop {} {\bibfield  {journal} {\bibinfo  {journal} {Physical Review
  B}\ }\textbf {\bibinfo {volume} {77}},\ \bibinfo {pages} {113409} (\bibinfo
  {year} {2008})}\BibitemShut {NoStop}%
\bibitem [{\citenamefont {Kim}\ and\ \citenamefont
  {Kim}(2008)}]{kim2008prediction}%
  \BibitemOpen
  \bibfield  {author} {\bibinfo {author} {\bibfnamefont {W.~Y.}\ \bibnamefont
  {Kim}}\ and\ \bibinfo {author} {\bibfnamefont {K.~S.}\ \bibnamefont {Kim}},\
  }\href@noop {} {\bibfield  {journal} {\bibinfo  {journal} {Nature
  Nanotechnology}\ }\textbf {\bibinfo {volume} {3}},\ \bibinfo {pages} {408}
  (\bibinfo {year} {2008})}\BibitemShut {NoStop}%
\bibitem [{\citenamefont {Mu{\~n}oz-Rojas}\ \emph {et~al.}(2009)\citenamefont
  {Mu{\~n}oz-Rojas}, \citenamefont {Fern{\'a}ndez-Rossier},\ and\ \citenamefont
  {Palacios}}]{munoz2009giant}%
  \BibitemOpen
  \bibfield  {author} {\bibinfo {author} {\bibfnamefont {F.}~\bibnamefont
  {Mu{\~n}oz-Rojas}}, \bibinfo {author} {\bibfnamefont {J.}~\bibnamefont
  {Fern{\'a}ndez-Rossier}}, \ and\ \bibinfo {author} {\bibfnamefont
  {J.}~\bibnamefont {Palacios}},\ }\href@noop {} {\bibfield  {journal}
  {\bibinfo  {journal} {Physical Review Letters}\ }\textbf {\bibinfo {volume}
  {102}},\ \bibinfo {pages} {136810} (\bibinfo {year} {2009})}\BibitemShut
  {NoStop}%
\bibitem [{\citenamefont {Zhang}\ \emph {et~al.}(2010)\citenamefont {Zhang},
  \citenamefont {Jiang}, \citenamefont {Sun},\ and\ \citenamefont
  {Xie}}]{zhang2010spin}%
  \BibitemOpen
  \bibfield  {author} {\bibinfo {author} {\bibfnamefont {Y.-T.}\ \bibnamefont
  {Zhang}}, \bibinfo {author} {\bibfnamefont {H.}~\bibnamefont {Jiang}},
  \bibinfo {author} {\bibfnamefont {Q.-F.}\ \bibnamefont {Sun}}, \ and\
  \bibinfo {author} {\bibfnamefont {X.}~\bibnamefont {Xie}},\ }\href@noop {}
  {\bibfield  {journal} {\bibinfo  {journal} {Physical Review B}\ }\textbf
  {\bibinfo {volume} {81}},\ \bibinfo {pages} {165404} (\bibinfo {year}
  {2010})}\BibitemShut {NoStop}%
\bibitem [{\citenamefont {Lu}\ \emph {et~al.}(2011)\citenamefont {Lu},
  \citenamefont {Zhang}, \citenamefont {Shi}, \citenamefont {Wang},
  \citenamefont {Zheng}, \citenamefont {Zhang}, \citenamefont {Wang},
  \citenamefont {Tang},\ and\ \citenamefont {Sheng}}]{lu2011graphene}%
  \BibitemOpen
  \bibfield  {author} {\bibinfo {author} {\bibfnamefont {J.}~\bibnamefont
  {Lu}}, \bibinfo {author} {\bibfnamefont {H.}~\bibnamefont {Zhang}}, \bibinfo
  {author} {\bibfnamefont {W.}~\bibnamefont {Shi}}, \bibinfo {author}
  {\bibfnamefont {Z.}~\bibnamefont {Wang}}, \bibinfo {author} {\bibfnamefont
  {Y.}~\bibnamefont {Zheng}}, \bibinfo {author} {\bibfnamefont
  {T.}~\bibnamefont {Zhang}}, \bibinfo {author} {\bibfnamefont
  {N.}~\bibnamefont {Wang}}, \bibinfo {author} {\bibfnamefont {Z.}~\bibnamefont
  {Tang}}, \ and\ \bibinfo {author} {\bibfnamefont {P.}~\bibnamefont {Sheng}},\
  }\href@noop {} {\bibfield  {journal} {\bibinfo  {journal} {Nano Letters}\
  }\textbf {\bibinfo {volume} {11}},\ \bibinfo {pages} {2973} (\bibinfo {year}
  {2011})}\BibitemShut {NoStop}%
\bibitem [{\citenamefont {Bai}\ \emph {et~al.}(2010)\citenamefont {Bai},
  \citenamefont {Cheng}, \citenamefont {Xiu}, \citenamefont {Liao},
  \citenamefont {Wang}, \citenamefont {Shailos}, \citenamefont {Wang},
  \citenamefont {Huang},\ and\ \citenamefont {Duan}}]{bai2010very}%
  \BibitemOpen
  \bibfield  {author} {\bibinfo {author} {\bibfnamefont {J.}~\bibnamefont
  {Bai}}, \bibinfo {author} {\bibfnamefont {R.}~\bibnamefont {Cheng}}, \bibinfo
  {author} {\bibfnamefont {F.}~\bibnamefont {Xiu}}, \bibinfo {author}
  {\bibfnamefont {L.}~\bibnamefont {Liao}}, \bibinfo {author} {\bibfnamefont
  {M.}~\bibnamefont {Wang}}, \bibinfo {author} {\bibfnamefont {A.}~\bibnamefont
  {Shailos}}, \bibinfo {author} {\bibfnamefont {K.~L.}\ \bibnamefont {Wang}},
  \bibinfo {author} {\bibfnamefont {Y.}~\bibnamefont {Huang}}, \ and\ \bibinfo
  {author} {\bibfnamefont {X.}~\bibnamefont {Duan}},\ }\href@noop {} {\bibfield
   {journal} {\bibinfo  {journal} {Nature Nanotechnology}\ }\textbf {\bibinfo
  {volume} {5}},\ \bibinfo {pages} {655} (\bibinfo {year} {2010})}\BibitemShut
  {NoStop}%
\bibitem [{\citenamefont {Friedman}\ \emph {et~al.}(2010)\citenamefont
  {Friedman}, \citenamefont {Tedesco}, \citenamefont {Campbell}, \citenamefont
  {Culbertson}, \citenamefont {Aifer}, \citenamefont {Perkins}, \citenamefont
  {Myers-Ward}, \citenamefont {Hite}, \citenamefont {Eddy~Jr}, \citenamefont
  {Jernigan} \emph {et~al.}}]{friedman2010quantum}%
  \BibitemOpen
  \bibfield  {author} {\bibinfo {author} {\bibfnamefont {A.~L.}\ \bibnamefont
  {Friedman}}, \bibinfo {author} {\bibfnamefont {J.~L.}\ \bibnamefont
  {Tedesco}}, \bibinfo {author} {\bibfnamefont {P.~M.}\ \bibnamefont
  {Campbell}}, \bibinfo {author} {\bibfnamefont {J.~C.}\ \bibnamefont
  {Culbertson}}, \bibinfo {author} {\bibfnamefont {E.}~\bibnamefont {Aifer}},
  \bibinfo {author} {\bibfnamefont {F.~K.}\ \bibnamefont {Perkins}}, \bibinfo
  {author} {\bibfnamefont {R.~L.}\ \bibnamefont {Myers-Ward}}, \bibinfo
  {author} {\bibfnamefont {J.~K.}\ \bibnamefont {Hite}}, \bibinfo {author}
  {\bibfnamefont {C.~R.}\ \bibnamefont {Eddy~Jr}}, \bibinfo {author}
  {\bibfnamefont {G.~G.}\ \bibnamefont {Jernigan}},  \emph {et~al.},\
  }\href@noop {} {\bibfield  {journal} {\bibinfo  {journal} {Nano Letters}\
  }\textbf {\bibinfo {volume} {10}},\ \bibinfo {pages} {3962} (\bibinfo {year}
  {2010})}\BibitemShut {NoStop}%
\bibitem [{\citenamefont {Liao}\ \emph {et~al.}(2012)\citenamefont {Liao},
  \citenamefont {Wu}, \citenamefont {Kumar}, \citenamefont {Duesberg},
  \citenamefont {Zhou}, \citenamefont {Cross}, \citenamefont {Shvets},\ and\
  \citenamefont {Yu}}]{liao2012large}%
  \BibitemOpen
  \bibfield  {author} {\bibinfo {author} {\bibfnamefont {Z.-M.}\ \bibnamefont
  {Liao}}, \bibinfo {author} {\bibfnamefont {H.-C.}\ \bibnamefont {Wu}},
  \bibinfo {author} {\bibfnamefont {S.}~\bibnamefont {Kumar}}, \bibinfo
  {author} {\bibfnamefont {G.~S.}\ \bibnamefont {Duesberg}}, \bibinfo {author}
  {\bibfnamefont {Y.-B.}\ \bibnamefont {Zhou}}, \bibinfo {author}
  {\bibfnamefont {G.~L.}\ \bibnamefont {Cross}}, \bibinfo {author}
  {\bibfnamefont {I.~V.}\ \bibnamefont {Shvets}}, \ and\ \bibinfo {author}
  {\bibfnamefont {D.-P.}\ \bibnamefont {Yu}},\ }\href@noop {} {\bibfield
  {journal} {\bibinfo  {journal} {Advanced Materials}\ }\textbf {\bibinfo
  {volume} {24}},\ \bibinfo {pages} {1862} (\bibinfo {year}
  {2012})}\BibitemShut {NoStop}%
\bibitem [{\citenamefont {Gopinadhan}\ \emph {et~al.}(2015)\citenamefont
  {Gopinadhan}, \citenamefont {Shin}, \citenamefont {Jalil}, \citenamefont
  {Venkatesan}, \citenamefont {Geim}, \citenamefont {Neto},\ and\ \citenamefont
  {Yang}}]{gopinadhan2015extremely}%
  \BibitemOpen
  \bibfield  {author} {\bibinfo {author} {\bibfnamefont {K.}~\bibnamefont
  {Gopinadhan}}, \bibinfo {author} {\bibfnamefont {Y.~J.}\ \bibnamefont
  {Shin}}, \bibinfo {author} {\bibfnamefont {R.}~\bibnamefont {Jalil}},
  \bibinfo {author} {\bibfnamefont {T.}~\bibnamefont {Venkatesan}}, \bibinfo
  {author} {\bibfnamefont {A.~K.}\ \bibnamefont {Geim}}, \bibinfo {author}
  {\bibfnamefont {A.~H.~C.}\ \bibnamefont {Neto}}, \ and\ \bibinfo {author}
  {\bibfnamefont {H.}~\bibnamefont {Yang}},\ }\href@noop {} {\bibfield
  {journal} {\bibinfo  {journal} {Nature Communications}\ }\textbf {\bibinfo
  {volume} {6}} (\bibinfo {year} {2015})}\BibitemShut {NoStop}%
\bibitem [{\citenamefont {Kisslinger}\ \emph {et~al.}(2015)\citenamefont
  {Kisslinger}, \citenamefont {Ott}, \citenamefont {Heide}, \citenamefont
  {Kampert}, \citenamefont {Butz}, \citenamefont {Spiecker}, \citenamefont
  {Shallcross},\ and\ \citenamefont {Weber}}]{kisslinger2015linear}%
  \BibitemOpen
  \bibfield  {author} {\bibinfo {author} {\bibfnamefont {F.}~\bibnamefont
  {Kisslinger}}, \bibinfo {author} {\bibfnamefont {C.}~\bibnamefont {Ott}},
  \bibinfo {author} {\bibfnamefont {C.}~\bibnamefont {Heide}}, \bibinfo
  {author} {\bibfnamefont {E.}~\bibnamefont {Kampert}}, \bibinfo {author}
  {\bibfnamefont {B.}~\bibnamefont {Butz}}, \bibinfo {author} {\bibfnamefont
  {E.}~\bibnamefont {Spiecker}}, \bibinfo {author} {\bibfnamefont
  {S.}~\bibnamefont {Shallcross}}, \ and\ \bibinfo {author} {\bibfnamefont
  {H.~B.}\ \bibnamefont {Weber}},\ }\href@noop {} {\bibfield  {journal}
  {\bibinfo  {journal} {Nature Physics}\ }\textbf {\bibinfo {volume} {11}},\
  \bibinfo {pages} {650} (\bibinfo {year} {2015})}\BibitemShut {NoStop}%
\bibitem [{\citenamefont {Zhai}\ and\ \citenamefont
  {Wang}(2016)}]{zhai2016atomistic}%
  \BibitemOpen
  \bibfield  {author} {\bibinfo {author} {\bibfnamefont {M.-X.}\ \bibnamefont
  {Zhai}}\ and\ \bibinfo {author} {\bibfnamefont {X.-F.}\ \bibnamefont
  {Wang}},\ }\href@noop {} {\bibfield  {journal} {\bibinfo  {journal}
  {Scientific Reports}\ }\textbf {\bibinfo {volume} {6}} (\bibinfo {year}
  {2016})}\BibitemShut {NoStop}%
\bibitem [{\citenamefont {Wu}\ \emph {et~al.}(2017)\citenamefont {Wu},
  \citenamefont {Chaika}, \citenamefont {Hsu}, \citenamefont {Huang},
  \citenamefont {Abid}, \citenamefont {Abid}, \citenamefont {Aristov},
  \citenamefont {Molodtsova}, \citenamefont {Babenkov}, \citenamefont {Niu}
  \emph {et~al.}}]{wu2017large}%
  \BibitemOpen
  \bibfield  {author} {\bibinfo {author} {\bibfnamefont {H.-C.}\ \bibnamefont
  {Wu}}, \bibinfo {author} {\bibfnamefont {A.~N.}\ \bibnamefont {Chaika}},
  \bibinfo {author} {\bibfnamefont {M.-C.}\ \bibnamefont {Hsu}}, \bibinfo
  {author} {\bibfnamefont {T.-W.}\ \bibnamefont {Huang}}, \bibinfo {author}
  {\bibfnamefont {M.}~\bibnamefont {Abid}}, \bibinfo {author} {\bibfnamefont
  {M.}~\bibnamefont {Abid}}, \bibinfo {author} {\bibfnamefont {V.~Y.}\
  \bibnamefont {Aristov}}, \bibinfo {author} {\bibfnamefont {O.~V.}\
  \bibnamefont {Molodtsova}}, \bibinfo {author} {\bibfnamefont {S.~V.}\
  \bibnamefont {Babenkov}}, \bibinfo {author} {\bibfnamefont {Y.}~\bibnamefont
  {Niu}},  \emph {et~al.},\ }\href@noop {} {\bibfield  {journal} {\bibinfo
  {journal} {Nature Communications}\ }\textbf {\bibinfo {volume} {8}} (\bibinfo
  {year} {2017})}\BibitemShut {NoStop}%
\bibitem [{\citenamefont {El-Ahmar}\ \emph {et~al.}(2017)\citenamefont
  {El-Ahmar}, \citenamefont {Koczorowski}, \citenamefont {Po{\'z}niak},
  \citenamefont {Ku{\'s}wik}, \citenamefont {Strupi{\'n}ski},\ and\
  \citenamefont {Czajka}}]{el2017graphene}%
  \BibitemOpen
  \bibfield  {author} {\bibinfo {author} {\bibfnamefont {S.}~\bibnamefont
  {El-Ahmar}}, \bibinfo {author} {\bibfnamefont {W.}~\bibnamefont
  {Koczorowski}}, \bibinfo {author} {\bibfnamefont {A.}~\bibnamefont
  {Po{\'z}niak}}, \bibinfo {author} {\bibfnamefont {P.}~\bibnamefont
  {Ku{\'s}wik}}, \bibinfo {author} {\bibfnamefont {W.}~\bibnamefont
  {Strupi{\'n}ski}}, \ and\ \bibinfo {author} {\bibfnamefont {R.}~\bibnamefont
  {Czajka}},\ }\href@noop {} {\bibfield  {journal} {\bibinfo  {journal}
  {Applied Physics Letters}\ }\textbf {\bibinfo {volume} {110}},\ \bibinfo
  {pages} {043503} (\bibinfo {year} {2017})}\BibitemShut {NoStop}%
\bibitem [{\citenamefont {Here}()}]{note1}%
  \BibitemOpen
  \bibfield  {author} {\bibinfo {author} {\bibnamefont {Here}},\ }\href@noop {}
  {}\bibinfo {note} {$W$ is several times of $l$ to ensure that the edge effect
  is negligible.}\BibitemShut {Stop}%
\bibitem [{\citenamefont {Zollner}\ \emph {et~al.}(2016)\citenamefont
  {Zollner}, \citenamefont {Gmitra}, \citenamefont {Frank},\ and\ \citenamefont
  {Fabian}}]{zollner2016theory}%
  \BibitemOpen
  \bibfield  {author} {\bibinfo {author} {\bibfnamefont {K.}~\bibnamefont
  {Zollner}}, \bibinfo {author} {\bibfnamefont {M.}~\bibnamefont {Gmitra}},
  \bibinfo {author} {\bibfnamefont {T.}~\bibnamefont {Frank}}, \ and\ \bibinfo
  {author} {\bibfnamefont {J.}~\bibnamefont {Fabian}},\ }\href@noop {}
  {\bibfield  {journal} {\bibinfo  {journal} {Physical Review B}\ }\textbf
  {\bibinfo {volume} {94}},\ \bibinfo {pages} {155441} (\bibinfo {year}
  {2016})}\BibitemShut {NoStop}%
\bibitem [{\citenamefont {Song}\ and\ \citenamefont
  {Wu}(2013)}]{song2013ballistic}%
  \BibitemOpen
  \bibfield  {author} {\bibinfo {author} {\bibfnamefont {Y.}~\bibnamefont
  {Song}}\ and\ \bibinfo {author} {\bibfnamefont {H.-C.}\ \bibnamefont {Wu}},\
  }\href@noop {} {\bibfield  {journal} {\bibinfo  {journal} {Journal of
  Physics: Condensed Matter}\ }\textbf {\bibinfo {volume} {25}},\ \bibinfo
  {pages} {355301} (\bibinfo {year} {2013})}\BibitemShut {NoStop}%
\bibitem [{\citenamefont {Song}\ \emph {et~al.}(2012)\citenamefont {Song},
  \citenamefont {Wu},\ and\ \citenamefont {Guo}}]{song2012giant}%
  \BibitemOpen
  \bibfield  {author} {\bibinfo {author} {\bibfnamefont {Y.}~\bibnamefont
  {Song}}, \bibinfo {author} {\bibfnamefont {H.-C.}\ \bibnamefont {Wu}}, \ and\
  \bibinfo {author} {\bibfnamefont {Y.}~\bibnamefont {Guo}},\ }\href@noop {}
  {\bibfield  {journal} {\bibinfo  {journal} {Applied Physics Letters}\
  }\textbf {\bibinfo {volume} {100}},\ \bibinfo {pages} {253116} (\bibinfo
  {year} {2012})}\BibitemShut {NoStop}%
\bibitem [{\citenamefont {Song}\ \emph {et~al.}(2013)\citenamefont {Song},
  \citenamefont {Zhai},\ and\ \citenamefont {Guo}}]{song2013generation}%
  \BibitemOpen
  \bibfield  {author} {\bibinfo {author} {\bibfnamefont {Y.}~\bibnamefont
  {Song}}, \bibinfo {author} {\bibfnamefont {F.}~\bibnamefont {Zhai}}, \ and\
  \bibinfo {author} {\bibfnamefont {Y.}~\bibnamefont {Guo}},\ }\href@noop {}
  {\bibfield  {journal} {\bibinfo  {journal} {Applied Physics Letters}\
  }\textbf {\bibinfo {volume} {103}},\ \bibinfo {pages} {183111} (\bibinfo
  {year} {2013})}\BibitemShut {NoStop}%
\bibitem [{\citenamefont {Born}\ and\ \citenamefont
  {Wolf}(1980)}]{born1980principles}%
  \BibitemOpen
  \bibfield  {author} {\bibinfo {author} {\bibfnamefont {M.}~\bibnamefont
  {Born}}\ and\ \bibinfo {author} {\bibfnamefont {E.}~\bibnamefont {Wolf}},\
  }\href@noop {} {\emph {\bibinfo {title} {Principles of optics:
  electromagnetic theory of propagation, interference and diffraction of
  light}}}\ (\bibinfo  {publisher} {Elsevier},\ \bibinfo {year}
  {1980})\BibitemShut {NoStop}%
\bibitem [{\citenamefont {Morozov}\ \emph {et~al.}(2008)\citenamefont
  {Morozov}, \citenamefont {Novoselov}, \citenamefont {Katsnelson},
  \citenamefont {Schedin}, \citenamefont {Elias}, \citenamefont {Jaszczak},\
  and\ \citenamefont {Geim}}]{morozov2008giant}%
  \BibitemOpen
  \bibfield  {author} {\bibinfo {author} {\bibfnamefont {S.}~\bibnamefont
  {Morozov}}, \bibinfo {author} {\bibfnamefont {K.}~\bibnamefont {Novoselov}},
  \bibinfo {author} {\bibfnamefont {M.}~\bibnamefont {Katsnelson}}, \bibinfo
  {author} {\bibfnamefont {F.}~\bibnamefont {Schedin}}, \bibinfo {author}
  {\bibfnamefont {D.}~\bibnamefont {Elias}}, \bibinfo {author} {\bibfnamefont
  {J.~A.}\ \bibnamefont {Jaszczak}}, \ and\ \bibinfo {author} {\bibfnamefont
  {A.}~\bibnamefont {Geim}},\ }\href@noop {} {\bibfield  {journal} {\bibinfo
  {journal} {Physical Review Letters}\ }\textbf {\bibinfo {volume} {100}},\
  \bibinfo {pages} {016602} (\bibinfo {year} {2008})}\BibitemShut {NoStop}%
\bibitem [{\citenamefont {Chen}\ \emph {et~al.}(2008)\citenamefont {Chen},
  \citenamefont {Jang}, \citenamefont {Xiao}, \citenamefont {Ishigami},\ and\
  \citenamefont {Fuhrer}}]{chen2008intrinsic}%
  \BibitemOpen
  \bibfield  {author} {\bibinfo {author} {\bibfnamefont {J.-H.}\ \bibnamefont
  {Chen}}, \bibinfo {author} {\bibfnamefont {C.}~\bibnamefont {Jang}}, \bibinfo
  {author} {\bibfnamefont {S.}~\bibnamefont {Xiao}}, \bibinfo {author}
  {\bibfnamefont {M.}~\bibnamefont {Ishigami}}, \ and\ \bibinfo {author}
  {\bibfnamefont {M.~S.}\ \bibnamefont {Fuhrer}},\ }\href@noop {} {\bibfield
  {journal} {\bibinfo  {journal} {Nature Nanotechnology}\ }\textbf {\bibinfo
  {volume} {3}},\ \bibinfo {pages} {206} (\bibinfo {year} {2008})}\BibitemShut
  {NoStop}%
\bibitem [{\citenamefont {B{\"u}ttiker}\ \emph {et~al.}(1985)\citenamefont
  {B{\"u}ttiker}, \citenamefont {Imry}, \citenamefont {Landauer},\ and\
  \citenamefont {Pinhas}}]{buttiker1985generalized}%
  \BibitemOpen
  \bibfield  {author} {\bibinfo {author} {\bibfnamefont {M.}~\bibnamefont
  {B{\"u}ttiker}}, \bibinfo {author} {\bibfnamefont {Y.}~\bibnamefont {Imry}},
  \bibinfo {author} {\bibfnamefont {R.}~\bibnamefont {Landauer}}, \ and\
  \bibinfo {author} {\bibfnamefont {S.}~\bibnamefont {Pinhas}},\ }\href@noop {}
  {\bibfield  {journal} {\bibinfo  {journal} {Physical Review B}\ }\textbf
  {\bibinfo {volume} {31}},\ \bibinfo {pages} {6207} (\bibinfo {year}
  {1985})}\BibitemShut {NoStop}%
\bibitem [{not()}]{note3}%
  \BibitemOpen
  \href@noop {} {}\bibinfo {note} {Note, this value would be smaller when taking
  into account the scattering.}\BibitemShut {Stop}%
\bibitem [{\citenamefont {Solin}\ \emph {et~al.}(2000)\citenamefont {Solin},
  \citenamefont {Thio}, \citenamefont {Hines},\ and\ \citenamefont
  {Heremans}}]{solin2000enhanced}%
  \BibitemOpen
  \bibfield  {author} {\bibinfo {author} {\bibfnamefont {S.}~\bibnamefont
  {Solin}}, \bibinfo {author} {\bibfnamefont {T.}~\bibnamefont {Thio}},
  \bibinfo {author} {\bibfnamefont {D.}~\bibnamefont {Hines}}, \ and\ \bibinfo
  {author} {\bibfnamefont {J.}~\bibnamefont {Heremans}},\ }\href@noop {}
  {\bibfield  {journal} {\bibinfo  {journal} {Science}\ }\textbf {\bibinfo
  {volume} {289}},\ \bibinfo {pages} {1530} (\bibinfo {year}
  {2000})}\BibitemShut {NoStop}%
\bibitem [{\citenamefont {Hewett}\ and\ \citenamefont
  {Kusmartsev}(2010)}]{hewett2010geometrically}%
  \BibitemOpen
  \bibfield  {author} {\bibinfo {author} {\bibfnamefont {T.~H.}\ \bibnamefont
  {Hewett}}\ and\ \bibinfo {author} {\bibfnamefont {F.}~\bibnamefont
  {Kusmartsev}},\ }\href@noop {} {\bibfield  {journal} {\bibinfo  {journal}
  {Physical Review B}\ }\textbf {\bibinfo {volume} {82}},\ \bibinfo {pages}
  {212404} (\bibinfo {year} {2010})}\BibitemShut {NoStop}%
\bibitem [{\citenamefont {Giovannetti}\ \emph {et~al.}(2008)\citenamefont
  {Giovannetti}, \citenamefont {Khomyakov}, \citenamefont {Brocks},
  \citenamefont {Karpan}, \citenamefont {Van~den Brink},\ and\ \citenamefont
  {Kelly}}]{giovannetti2008doping}%
  \BibitemOpen
  \bibfield  {author} {\bibinfo {author} {\bibfnamefont {G.}~\bibnamefont
  {Giovannetti}}, \bibinfo {author} {\bibfnamefont {P.}~\bibnamefont
  {Khomyakov}}, \bibinfo {author} {\bibfnamefont {G.}~\bibnamefont {Brocks}},
  \bibinfo {author} {\bibfnamefont {V.~v.}\ \bibnamefont {Karpan}}, \bibinfo
  {author} {\bibfnamefont {J.}~\bibnamefont {Van~den Brink}}, \ and\ \bibinfo
  {author} {\bibfnamefont {P.}~\bibnamefont {Kelly}},\ }\href@noop {}
  {\bibfield  {journal} {\bibinfo  {journal} {Physical Review Letters}\
  }\textbf {\bibinfo {volume} {101}},\ \bibinfo {pages} {026803} (\bibinfo
  {year} {2008})}\BibitemShut {NoStop}%
\end{thebibliography}

%

\end{document}